\documentclass[aps,pra,twocolumn,10pt,superscriptaddress,showpacs,nofootinbib
]{revtex4-1}
\usepackage{hyperref,xcolor}

\newcommand{\QUOTE}[3]{\begin{quote} \emph{``#1"} \\ \rule{0pt}{0pt} \hfill (#3) \end{quote}}

\begin{document}

\title{Nothing works the first time: An expert experimental physics epistemology}

\author{Dimitri R. Dounas-Frazer}
\affiliation{Department of Physics, University of Colorado Boulder, Boulder, CO 80309, USA}

\author{H. J. Lewandowski}
\affiliation{Department of Physics, University of Colorado Boulder, Boulder, CO 80309, USA}
\affiliation{JILA, National Institute of Standards and Technology and University of Colorado Boulder, Boulder, CO 80309, USA}


\date{\today}

\begin{abstract}
{The ability to troubleshoot is an important learning outcome for undergraduate physics laboratory courses. To better understand the role of troubleshooting in electronics laboratory courses, we interviewed 20 electronics instructors from multiple institution types about their beliefs and teaching practices related to troubleshooting. In these interviews, instructors articulated the idea that \emph{nothing works the first time} in multiple contexts pertaining to troubleshooting. We argue that this idea is an expert epistemology and show how it informs instructors' beliefs that (i) students need to know how to troubleshoot, (ii) students should expect to troubleshoot, (iii) all circuit-building lab activities provide opportunities for students to troubleshoot, and (iv) students' ability to construct functional circuits can be a proxy for their ability to troubleshoot malfunctioning circuits. Moreover, we discuss implications for instruction and assessment of troubleshooting in electronics courses.}
\end{abstract}

\maketitle


\section{Introduction}

Troubleshooting is relevant to a broad range of technical professions~\cite{Perez1991,Schaafstal2000,Jonassen2006}, including experimental physics~\cite{Pollard2014}. In particular, the ability to troubleshoot is an important design-related learning goal for undergraduate physics laboratory courses~\cite{AAPT2015,Zwickl2012,Zwickl2013}. Electronics courses, which typically engage students in circuit-building lab activities, are particularly well suited to developing students' troubleshooting abilities: circuit components are cheap and easy to replace, and---as we unpack in the present work---the need to troubleshoot arises naturally since students often build circuits that don't initially work. However, despite the importance of troubleshooting in electronics courses, there is a dearth of research-based instructional materials related to troubleshooting in this context.

Previous work on troubleshooting circuits has focused on learners' actions and interactions, for example: the roles of model-based reasoning~\cite{Dounas-Frazer2015,Dounas-Frazer2016} and socially-mediated metacognition~\cite{VanDeBogart2015} in undergraduate students' approaches to troubleshooting a malfunctioning circuit, and expertise-related differences among high school students troubleshooting simulated circuits~\cite{VanGog2005b}. While recent work~\cite{Coppens2016} focused on electronics instructors' views about learning goals for electronics courses, we are unaware of work that focuses on instructors' perspectives on, or experiences with, troubleshooting instruction.

As is true for test development in other physics education contexts~\cite{Engelhardt2009,Wilcox2015}, understanding instructors' views and practices related to teaching and learning troubleshooting is an important step towards clarifying the need for, and objectives of, research-based assessments of students' ability to troubleshoot. To this end, we interviewed 20 electronics instructors from a wide variety of institution types across the United States about their approaches to teaching and assessing troubleshooting. We found that several instructors' approaches are informed by the belief that \emph{nothing works the first time}, that is, a given experimental system---including a circuit---will probably not perform as expected immediately after being constructed. Similar to other work on physics epistemologies~\cite{Hammer1994}, we characterized this idea based on interviewees' statements about physics, learning physics, teaching, or electronics courses more generally.

In this work, we elaborate on how instructors' belief that nothing works the first time influenced multiple facets of their instruction in electronics courses: because nothing works the first time, it follows that (i) students need to know how to troubleshoot, (ii) students should expect to troubleshoot, (iii) all circuit-building lab activities provide opportunities for students to troubleshoot, and (iv) students' ability to construct functional circuits is a proxy for their ability to troubleshoot malfunctioning circuits. We argue that the idea that nothing works the first time is an expert experimental physics epistemology that nevertheless has limitations: sometimes, students' circuits \emph{do} work the first time. In addition, we discuss implications for teaching and assessing troubleshooting.


\section{Methods and participants}

We conducted semi-structured interviews with electronics instructors to gain insight into their experiences with, and perspectives on, the role of troubleshooting in electronics courses. Our interview protocol consisted of 29 questions: 10 focused on departmental and course context, 2 on participants' teaching history, 4 on the value of troubleshooting and its connection to the purpose of electronics, 12 on teaching and assessment practices related to troubleshooting, and 1 final question about participants' race, ethnicity, and gender. Most deviations from the protocol were instances where the interviewer asked a participant to clarify or elaborate on an idea.

One goal of this study was to interview instructors teaching upper-division electronics courses at a variety of institution types. To accomplish this goal, we generated a database of electronics instructors and used the Carnegie classification system~\cite{Carnegie2015} to characterize the institutions with which they were affiliated. Our database was initially populated with instructors from our own professional networks as well as members of the Advanced Laboratory Physics Association~\cite{ALPhA} who participated in conference sessions and workshops related to electronics at the 2015 Conference on Laboratory Instruction Beyond the First Year. In addition, we perused websites for physics departments at Minority-Serving Institutions and Women's Colleges in order to identify whether those departments offered an electronics course. If so, we added the corresponding instructor to our database.

During Fall 2015, we solicited participation from 47 instructors via email. In total, 20 instructors participated in our study: 15 identified as white or Caucasian alone, 2 identified as mixed race (1 white and Black, 1 Caucasian with Cherokee and African background), and 1 each identified as Asian Indian, Mexican American, and Persian; 14 identified as male and 6 as female. In order to maintain anonymity of research participants, we do not report intersections of race or ethnicity and gender.

We interviewed electronics instructors from 18 distinct institutions: 12 public and 6 private not-for-profit institutions; 9 Predominantly White Institutions, 6 Hispanic-Serving Institutions, 2 Women's Colleges, 1 Historically Black University, and 1 Tribal College. One institution was classified as both Predominantly White and a Women's College. In terms of size and selectivity, instructors from small, medium, and large institutions as well as from inclusive, selective, and more selective institutions were about equally represented in our data set. Three institutions were two-year colleges, 5 were four-year institutions, 8 were Master's-granting institutions, and 3 were universities with doctoral physics programs.

Interviews were conducted via videoconference or in person. Audio data were recorded for each interview. Each interview lasted about 35--55 minutes, for a cumulative total of about 15 hours of audio data. One of us (D.R.D.F.) conducted and transcribed all interviews. The transcripts were the data that we analyzed. Six themes informed our research goals, interview protocol, and analyses: the purpose of electronics courses, the value of troubleshooting, the definition of troubleshooting, characteristics of proficient troubleshooting, methods of teaching troubleshooting, and methods of assessing troubleshooting. These themes served as an \emph{a priori} coding scheme: for each theme, one of us (D.R.D.F.) read through each transcript and identified related ideas that both authors discussed and collaboratively grouped into subthemes. Each transcript was read a total of six times. Here we report on one subtheme---nothing works the first time---that emerged across multiple themes.


\section{Results and interpretation}

We show that instructors' belief that nothing works the first time underlies four ideas about instruction:  students need to know how to troubleshoot, students should expect to troubleshoot, all activities are opportunities to troubleshoot, and activity completion implies successful troubleshooting.


Instructors indicated that, since nothing works the first time, it is important for students to develop the ability to troubleshoot:
\QUOTE{
I think [the ability to troubleshoot is] the most important thing [students] take away. Anything they're gonna do in the future, almost nothing in the real world works the first time. If it did, it was done a hundred years ago. Everything has issues, nothing works the way it's supposed to. And that's where you need a thinking, competent scientist or engineer to make your way through.}{I19}{Evergreen}
\QUOTE{I think it's important to know how to troubleshoot. It's rare that you---especially for complicated circuits---you whip it up and it works right away. No! Any experiment does not work right away. So you have to know about how to make it work.}{I14}{Maple}
\QUOTE{I think [the ability to troubleshoot is] very important. These students mostly come in thinking that when they draw a design on paper, they're gonna put it together and it's gonna work. So it's a rude awakening to see that that's rarely the case. As the circuits get more complicated.}{I18}{Birch}
According to Evergreen, Maple, and Birch, ``almost nothing in the real world works the first time," ``[a]ny experiment does not work right away," and the circuits that students build in electronics courses are sufficiently complicated that it's ``rarely the case" that they work immediately after being built. As a result, troubleshooting is the work of a ``thinking, competent scientist or engineer;" it is something experimentalists ``have to know about." And the ability to troubleshoot, then, is a ``very important" skill that students should learn.


Birch also implied that inexperienced troubleshooters are marked by a lack of anticipation of the need to troubleshoot their circuit: they think that ``they're gonna put it together and it's gonna work." Birch went on to say,
\QUOTE{The main point of [the final project] is for [students] to see how difficult it is to do the troubleshooting and that that's not---and that it's gonna be very necessary because as soon as they start making circuits that involve more than one subsystem, you connect them together and find they don't work. [Students] should expect that.}{I18}{Birch}
%
%
According to Birch, students should view troubleshooting as ``necessary" and they should ``expect" that their circuits won't work, especially when building circuits comprised of multiple subsystems. Similarly, other instructors also drew connections between students' proficiency and their expectations about the need to troubleshoot:
\QUOTE{It's getting around this whole thing, `If I built this circuit the way the diagram says, it should work.' It's a binary thing. `I built it, it works.' You have to get over that. That's not the case. There are things that can go wrong. It's important to get over that hurdle.}{I12}{Yew}
\QUOTE{It seems like it is hard for [students] to let go of the, `If I put it together and follow instructions, it should work the first time.' You need to knock that down before you teach troubleshooting. \ldots\ [Interviewer: What kind of attitude or mentality do you try to switch people into instead?] `Of course it won't work the first time! Come on!'}{I07}{Larch}
Yew and Larch each articulated that one characteristic of inexperienced troubleshooters is the belief that following instructions while constructing a circuit will always result in a working circuit. This belief, according to the instructors, is a ``hurdle" that students need to ``get over;" it is something that instructors need to ``knock down." Consistent with Birch, Yew also suggested that students should expect that their circuit ``won't work the first time." Thus, the belief that nothing works the first time is connected to anticipation of the need to troubleshoot, both of which are perceived by instructors to be desirable characteristics of experienced students.


Instructors' belief that nothing works the first time also informs their belief that all electronics lab activities are opportunities to troubleshoot. For example, Pine said,
\QUOTE{Due to its nature, every electronics lab is specifically designed to increase a students' ability to troubleshoot. Nobody is able to step in and wire them correctly the first time.}{I11}{Pine}
Pine's statement is consistent with the sentiment that troubleshooting is a natural part of electronics labs. Other instructors shared this sentiment, though they expressed different perceptions about the intent behind activity design. When asked whether they have implemented lab activities specifically designed to develop students' troubleshooting skills, most instructors said they had not done so:
\QUOTE{No. Again because I feel like [the need to troubleshoot] comes up in the natural course of things and we're kind of limited on time.}{I10}{Dogwood}
\QUOTE{No. I have basically assigned projects and let the troubleshooting happen on its own. \ldots\  I've kind of been counting on it happening by accident.}{I03}{Tanoak}
\QUOTE{That was not ever the primary goal [of any lab activity]. \ldots\  In some ways, none of them are explicitly about troubleshooting, but it's built into most of them.}{I01}{Filbert}
Dogwood, Tanoak, and Filbert indicated that the need to troubleshoot is a ``natural" part of labs, it happens ``by accident," and it is ``built into" most lab activities. This reasoning is rooted in the belief, articulated by Pine, that ``[n]obody is able to step in and wire [their circuit] correctly the first time."


Many instructors reasoned that, because every lab activity requires students to troubleshoot, students' construction of a functional circuit can therefore be used as a proxy for demonstrating proficient troubleshooting ability:
\QUOTE{Obviously, the [students] who made their way through [the lab activity] fast and efficiently were good at [troubleshooting]. \ldots\ But do I know the details of why they're good? Whether they're meticulous or whether they naturally debugged, intuitively---I don't know.}{I19}{Evergreen}
\QUOTE{I have students that are good, but I don't know if they're just good at preventing errors or good at finding them. Probably they're good at both, would be my guess. The ones that really finish up lab way before everyone else just seem to have a knack for getting things working. \ldots\ But the best way [to tell if students are good at troubleshooting] is, `Can they get it done?'}{I08}{Willow}
\QUOTE{It becomes more difficult to think about how to assess [the ability to troubleshoot]. Would I say that students are graded on their ability to troubleshoot? No. \ldots\  The main portion of their grade is based on lab, so if they can troubleshoot enough to get their lab to work---usually I help them out---I wouldn't say it's assessed directly. You could get out of the course without troubleshooting a lot because you built your circuits correctly every time.}{I01}{Filbert}
Evergreen and Willow indicated that students who complete lab activities ``fast and efficiently" or ``way before everyone else" are good at troubleshooting. Similarly, Filbert described assessing students' ability to troubleshoot indirectly: students who successfully complete their lab activities ``can troubleshoot enough to get their lab to work." However, the idea that nothing works the first time is in conflict with the idea that ``meticulous" circuit construction can prevent errors, resulting in circuits that are built correctly and work with little or no troubleshooting. Evergreen, Willow, and Filbert all recognized that using the construction of functional circuits as an indirect measure of troubleshooting ability has limitations that stem from the fact that, sometimes, students' circuits \emph{do} work the first time.


\section{Discussion}

Our results suggest that the idea that nothing works the first time is an expert experimental physics epistemology. This idea was not only prevalent among instructors in our study---espoused by 14 out of 20 interviewees---it also informed their beliefs and practices about teaching and learning troubleshooting in electronics courses. In addition, instructors articulated that one characteristic of students' proficiency with troubleshooting is the anticipation of the need to troubleshoot, which is rooted in \emph{students'} belief that nothing works the first time. Beyond the context of electronics, instructors indicated that the idea that nothing works the first time applies broadly to ``the real world" and ``[a]ny experiment," indicating that this idea is an epistemological belief about the domain of experimental physics in general~\cite{Hammer1994}.

Based on our own experience doing experimental physics research and teaching upper-division electronics course, we share the view that most circuit-building lab activities naturally give rise to the need to troubleshoot. Consistent with this view, we highlight two implications for instruction. First, regarding improvement of instruction about troubleshooting, instructors and education researchers can supplement existing lab activities with explicit instruction about troubleshooting strategies.

Second, it is not universally true that nothing works the first time; students sometimes build circuits that function correctly without needing to be troubleshot. Thus, factors aside from troubleshooting can also contribute to student efficiency and successful completion of lab activities. Hence outcomes-based assessments that focus on students' ability to construct functional circuits are insufficient for determining whether students are proficient troubleshooters. Process-based assessments that focus on students' troubleshooting strategies may be more appropriate.



\section{Conclusion}

We reported results from interviews with 20 electronics instructors from 18 institutions. We found that the idea that nothing works the first time is an expert epistemology about experimental physics. This epistemology underlies four ideas about troubleshooting instruction: students need to know how to troubleshoot, students should expect to troubleshoot, all activities are opportunities to troubleshoot, and activity completion implies successful troubleshooting. Future work will elaborate on instructors' characterization of proficient troubleshooting. Ultimately, this work will inform the design of assessments of students' ability to troubleshoot electric circuits.

\acknowledgments We acknowledge Gina Quan, Kevin Van De Bogart, and MacKenzie Stetzer for providing valuable feedback. This material is based upon work supported by the NSF under Grants No. DUE-1323101 and PHY-1125844.

\bibliography{ts_database_260516-short}

\end{document}